\documentclass[twocolumn,prc,aps,superscriptaddress,showpacs]{revtex4}
\usepackage{graphicx}
\usepackage{amssymb}
\usepackage[latin2]{inputenc}
\usepackage[T1]{fontenc}
\begin{document}

\title{On the possibility to observe neutron dark decay in nuclei}

\date{\today}


\author{M.~Pf\"utzner}
\email{pfutzner@fuw.edu.pl}
\affiliation{Faculty of Physics, University of Warsaw, 02-093 Warsaw, Poland}
\author{K.~Riisager}
\affiliation{Department of Physics and Astronomy, Aarhus University, DK-8000 Aarhus C, Denmark}

\begin{abstract}
As proposed recently by Fornal and Grinstein, neutrons can undergo a dark matter decay mode
which was not observed before. Such a decay could explain the existing discrepancy between
two different methods of neutron lifetime measurements. If such neutron decay is possible, then it should occur also is nuclei with sufficiently low neutron binding energy.
We examine a few nuclear candidates for the dark neutron decay and we consider  possibilities of their experimental identification. In more detail we discuss the case
of $^{11}$Be which appears as the most promising nucleus for the observation of the
neutron dark decay.
\end{abstract}

\maketitle


The neutron lifetime remains one of the most remarkable open questions in fundamental physics \cite{RMP83,Abele08}. After decades of experimental struggle two different experimental methods appear to yield two lifetime values with the discrepancy between them on the level of
about 4.0$\sigma$. The first type of measurements, referred to as the \emph{bottle} essentially counts the number of neutrons in the container as a function of time. In the second method, called the \emph{beam}, one counts protons resulting from beta decay of known number of neutrons in the beam. On average, the latter method yields the longer neutron lifetime by about 1\% \cite{RMP83}. Strictly speaking, however, both method measure different observables. While the bottle method provides the total lifetime of the neutron, the beam method is sensitive to the partial neutron lifetime related to beta decay. The former value being smaller indicates that other decay channels may contribute to the total lifetime.
In a recent paper, Fornal and Grinstein \cite{Fornal} formulate a hypothesis that the neutron undergoes a yet unobserved decay mode involving dark sector particles in the final state.
If such a new dark decay channel occurs with the branching ratio of 1\%, it would explain the neutron lifetime puzzle. Fornal and Grinstein discuss the energy constrains for the dark decay and proceed by investigating a few theoretical scenarios for the postulated dark channel \cite{Fornal}. They observe that neutrons bound in nuclei can also decay by the dark channel, if allowed by energy conditions.

In this paper we examine the possibility of nuclear dark neutron decay. First, we recall the definitions, energy conditions, and specify our assumptions. Then, a few nuclear candidates will be presented and their experimental prospects will be briefly discussed. Finally, the most promising case will be described in more detail.

Let the sum of masses of particles in the final state of the free neutron dark decay be $m_X$. From the stability of $^{9}$Be, Fornal and Grinstein give the following condition for $m_X$ \cite{Fornal}:
\begin{equation}
  937.900 \, {\rm MeV} < m_X < 939.565 \,{\rm MeV} ,
\end{equation}
where the lower limit is the atomic mass difference of $^{9}$Be and $^{8}$Be, and the upper limit is the neutron mass $m_n$. However, $^{8}$Be is not bound and it promptly disintegrates into two $\alpha$ particles. In consequence, the lower limit for the $m_X$ is actually given by the mass difference of $^{9}$Be and two $\alpha$ particles, which is 93 keV larger that the limit in Eq.1 \cite{AME2012}.
When the decaying neutron is bound in the nucleus, the dark decay may occur only if $m_X$ value happens to be in the reduced energy range:
\begin{equation}
  937.993 \, {\rm MeV} < m_X < m_n - S_n ,
\end{equation}
where $S_n$ is the neutron separation energy for the initial nucleus. This range is
open (larger than zero) if $S_n < 1.572 \, {\rm MeV}$.
We note that the final state in this scenario in principle can be identified through
the emergence of a less massive daughter nucleus, allowing model independent tests
of the dark decay in a restricted mass range.

The two neutron lifetime values given in Ref.\cite{Fornal} point to a partial half-life for the free neutron dark decay of $T_{1/2}^{nX} = 67 707$~s. In the following we will assume that the dark decay products do not interact with nucleons and that the matrix element governing the neutron dark decay in the nuclear environment is the same as for the free neutron. This is likely to be a conservative assumption, as an example the matrix element for radiative decay will decrease with decreasing available energy.
It follows that the partial decay rate for the dark decay in a nucleus will differ from that of the free neutron only by the density of final states (phase-space) factor, which in general will be smaller than one. First, for the sake of rough and first approximation, we will neglect the phase-space factor and estimate the upper limit of the branching ratio for the dark decay in the nucleus by
\begin{equation}
  B_X = \frac{T_{1/2}}{T_{1/2}^{nX}} ,
\end{equation}
where $T_{1/2}$ is the measured half-life of the nucleus.
Another reduction of the dark decay rate may result from angular momentum consideration when the decay products have to take away non-zero orbital angular momentum.

We proceed by discussing a few nuclear candidates for the dark decay. All values of nuclear properties are taken from the NNDC compilation \cite{NNDC}. Apart from nuclei with a small one-neutron separation energy we should also consider nuclei that, due to the odd-even staggering, have a two-neutron separation energy lower than the one for a single neutron. The best candidates are the so-called halo nuclei, see tables 2 and 3 in \cite{Rii13} for an overview.

\vspace{3mm}

\textbf{$^{6}$He}.
The two-neutron separation energy is 975.45(5) keV whereas a transition to the broad ground state of the unbound one-neutron daughter $^{5}$He requires around 1.7 MeV. A dark decay could proceed to a final nuclear state with an alpha particle and a single neutron. Observation of a free neutron from $^{6}$He decay would, although difficult to do, be a unique signature.
The half-life of $^{6}$He, $T_{1/2} = 807$~ms, leads to the branching ratio upper limit of
$B_X = 1.2 \times 10^{-5}$.
However, the nuclear matrix element for the transition would be reduced and the allowed energy window is small. This makes the decay less competitive.

\textbf{$^{11}$Li}.
This case has been already pointed to by Fornal and Grinstein \cite{Fornal}.
The neutron separation energy in $^{11}$Li of $S_n = 396(13)$~keV is favorite for the search of the dark decay producing the unbound $^{10}$Li which would promptly decay to $^{9}$Li by emission of a neutron. 
So in fact the allowed energy window for the dark decay in this case is given by the two-neutron separation energy equal to 369.3(6)~keV. 
However, the short half-life of $^{11}$Li, $T_{1/2} = 8.75$~ms,
yields the branching ratio
$B_X = 1.3 \times 10^{-7}$. Unfortunately, this value is by three orders of magnitude smaller than the branching ratio for the $\beta$-delayed deuteron emission from $^{11}$Li, $B_d = 1.30(13) \times 10^{-4}$ \cite{Raabe}, which leads to the same final nucleus $^{9}$Li.
In addition, the $\beta$-delayed proton emission of $^{11}$Li, leading to $^{10}$Li, is energetically allowed, although not yet observed. Experimental disproval of this channel at the level of $10^{-7}$ or lower, will be extremely difficult. Therefore, $^{11}$Li can not be considered as a practical candidate for the nuclear dark decay.

\textbf{$^{11}$Be}.
The neutron separation energy in this case is $S_n = 501.6(3)$~keV and the half-life is
$T_{1/2} = 13.76(7)$~s. Then, the branching ratio for the dark decay should be smaller
than $B_X = 2.0 \times 10^{-4}$. The dark decay of $^{11}$Be leads to the very long-lived $^{10}$Be ($T_{1/2} = 1.51(4)$~My). The appearance of $^{10}$Be in a sample originally containing only $^{11}$Be can be demonstrated by the accelerator mass spectrometry (AMS) method. In fact, such a measurement has been already done for $^{11}$Be in a search for $\beta$-delayed proton emission from this nucleus with a positive result \cite{Riisager}. Thus, $^{11}$Be appears as a good candidate for the observation of the neutron dark decay. We will analyse this case in more detail below.

\textbf{$^{15}$C}.
The rather large neutron separation energy, $S_n = 1218.1(8)$~keV, leaves a small energy region for probing the dark decay. In turn, it has a favorite half-life, $T_{1/2} = 2.449(5)$~s, yielding $B_X = 3.6 \times 10^{-5}$ which is not beyond experimental reach. Moreover, $\beta$-delayed proton emission is energetically not allowed in this case and the product of the dark decay is the long lived $^{14}$C ($T_{1/2} = 5.70(3)$~ky). The highly specialized AMS techniques have been developed to determine minute amounts of $^{14}$C with the large accuracy for the purpose of dating. Therefore, there should be not difficult to establish the appearance of $^{14}$C from a sample containing originally only $^{15}$C.

\textbf{$^{17}$C}.
The neutron separation energy is $S_n = 734(18)$~keV and the half-life $T_{1/2} = 193(5)$~ms.
The dark branching ratio $B_X = 3 \times 10^{-6}$ results. In addition to this very small value, a good signature of the dark decay is missing. In this decay $^{16}$C is produced which decays in 99\% by $\beta$-delayed neutron emission to the stable $^{15}$N. However, the same final nucleus can be populated in the possible $\beta$-delayed two-neutron emission from $^{17}$C. In consequence, $^{17}$C is not a practical choice to search for the dark decay.

\vspace{3mm}

There are several heavier nuclei fulfilling the energy condition $S_n < 1.572 \, {\rm MeV}$. All of them, however, have half-lives of the order of 10~ms. In addition, these are very neutron-rich isotopes, which are produced in minute amounts in the present day radioactive beam facilities. As a result, they do not represent practical candidates for the search of the neutron dark decay. Finally, from our survey $^{11}$Be emerges as the best candidate for this new, exotic decay channel.

$^{11}$Be attracts attention since long as a one-neutron halo nucleus in which several $\beta$-delayed particle emission channels are open. In particular, $\beta$-delayed proton
emission is possible with the $Q_{\beta p}$ value of 280~keV. Observation of this decay branch offers an interesting probe of the neutron-halo wave function which is expected to exhibit the single-particle behaviour. In a simplistic view, the halo neutron undergoes beta decay
into a proton which flies away. Due to the low energy value, however, this decay mode is strongly suppressed and was never observed directly. An important step in the studies of $^{11}$Be was taken recently by Riisager et al, who collected a sample of $^{11}$Be atoms
at the CERN-ISOLDE facility and subsequently measured it using the AMS technique at the VERA facility in Vienna \cite{Riisager}. They did identify atoms of $^{10}$Be and claimed the
observation of $\beta$p channel with the branching of $B_p = 8.3(9) \times 10^{-6}$ \cite{Riisager}. However, as discussed above, the appearance of $^{10}$Be atoms in such
experiment may result also from the dark decay of $^{11}$Be. To shed light on this alternative,
an additional measurement has to be done aiming at the direct observation of protons emitted in the $\beta$p decay of $^{11}$Be. Such experiment is planned in the near future at the CERN-ISOLDE laboratory.

The appearance of $^{10}$Be determined by Riisager et al. is by two orders of magnitude smaller than the rough estimate $B_X$ given above. The real branching ratio for the dark decay may be much smaller, for example due to the phase-space factor which depends on the mass $m_X$.
We investigate a possible influence of this factor in a simple model. We assume the decay scenario in which the neutron decays into two particles, one of mass $m_X$ and the second of negligible mass or massless. It can be shown that when a decay energy $Q$ is much smaller
than the mass of decaying body, the density of states, and thus the decay rate, is proportional to $Q^2$. Then, assuming that only one least-bound neutron can decay, the better approximation of the nuclear branching ratio will be obtained by multiplying the $B_X$ (Eq.3) by the factor
\begin{equation}
  f = \frac{(m_n - m_X - S_n)^2}{(m_n - m_X)^2} \, .
\end{equation}
The resulting branching ratio for $^{11}$Be, as a function of mass $m_X$ is shown in Fig. 1.
\begin{figure}[th]
\begin{center}
    \includegraphics[width=1.0\linewidth]{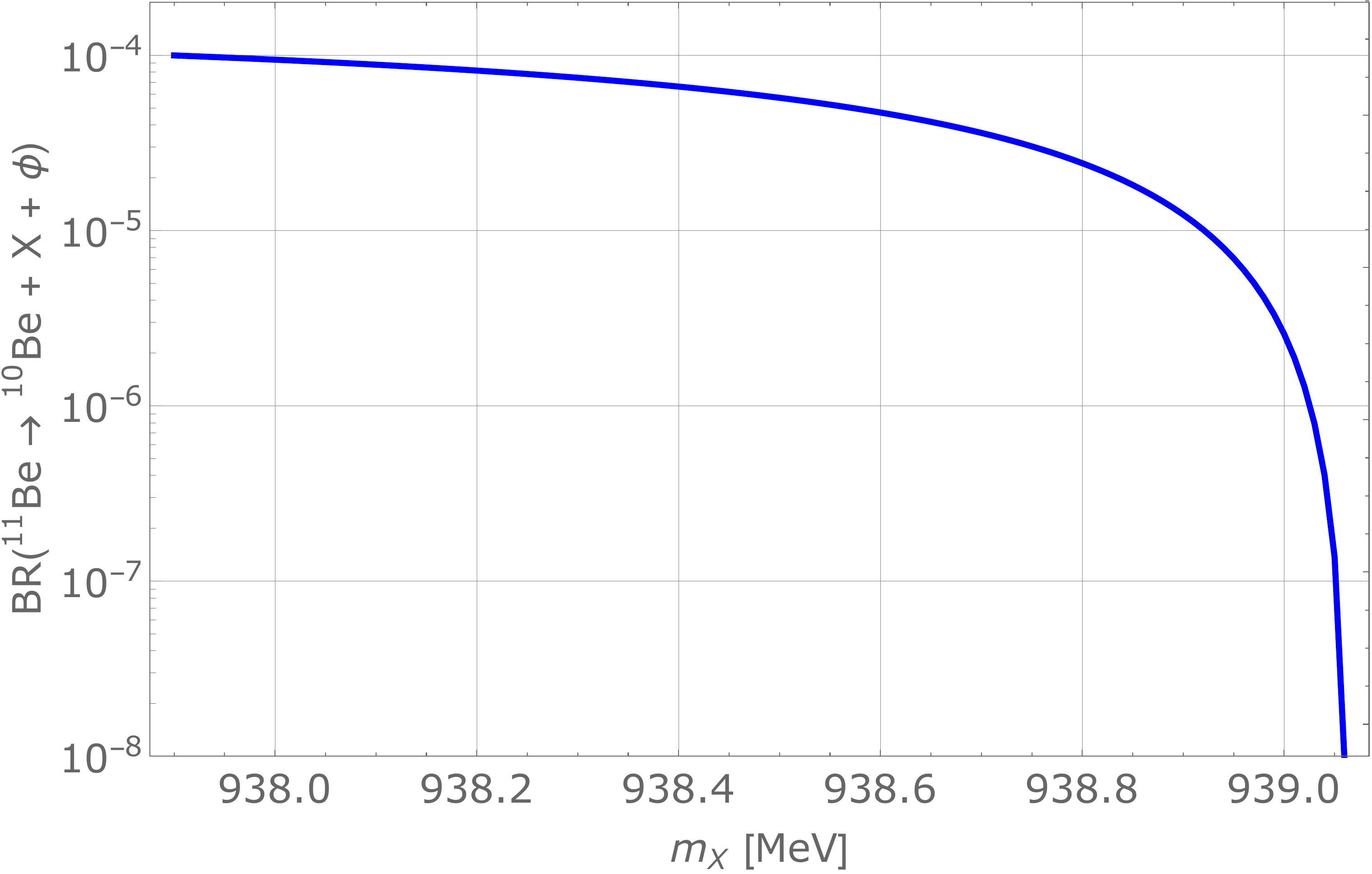}
    \caption{The estimated branching ratio for the dark decay of $^{11}$Be as a function of mass $m_X$.}
\end{center}
\end{figure}

The current data on $^{11}$Be decays into $^{10}$Be would correspond to a 95\% confidence  upper limit on the branching ratio of $10^{-5}$. That number can be converted to a lower limit of 938.9 MeV on the mass of the X particle. Other assumptions concerning the final state, however, would yield different phase-space energy dependence and different limits on postulated new particles.

We note that the limit obtainable from non-observation of dark decays in nuclei is relevant also for other decay modes by which a neutron may disappear, not only the decay branches considered in Ref. \cite{Fornal}. Their proposal has triggered much recent activity including a strong limit on the radiative dark decay mode \cite{Tang18}, the observation that the resolution of the neutron lifetime discrepancy could be coupled to a similar discrepancy for the axial coupling constant \cite{Czar18}, and a recent paper exploring also decays to mirror neutrons \cite{Sere18}. It was also shown that the existence of neutron stars with masses above
$0.7 \, M_{\bigodot}$ puts strong constrains on possible dark particles having a mass close to that of the neutron \cite{McKeen18,Baym18,Motta18}. The latter argument, however, involves various assumptions concerning interaction properties of such particles. We stress that investigation of nuclear decays discussed in this paper offers a model independent way to search for possible new decay channels of the neutron.

As a final remark we note that if indeed the neutron can decay to unknown, very weekly
interacting, or sterile particles, the only evidence for such a process in experiments
with free neutrons is the disappearance of neutrons. In contrast, the same process occurring
in a nucleus is manifested by the appearance of a different nucleus - a lighter isotope of the initial one. The observation of the latter phenomenon should be easier and in principle it could be demonstrated in a convincing way.

We would like to thank Bogdan Fornal, Dario Vretenar, Hans Fynbo, and Jonas Refsgaard for helpful discussions.


\begin{thebibliography}{999}
%
\bibitem{RMP83}
F.E. Wietfeld and G.L. Greene, Rev. Mod. Phys. \textbf{83}, 1173 (2011).
%
\bibitem{Abele08}
H. Abele, Prog. Part. Nucl. Phys. \textbf{60}, 1 (2008).
%
\bibitem{Fornal}
Bartosz Fornal and Benjamin Grinstein, arXiv:1801.01124v2 [hep-ph], (2018).
%
\bibitem{AME2012}
M. Wang et al., Chinese Phys. C \textbf{36}, 1603 (2012).
%
\bibitem{NNDC} http://www.nndc.bnl.gov/ensdf/
%
\bibitem{Rii13}
K. Riisager, Phys. Scr. \textbf{T152}, 014001 (2013).
%
\bibitem{Raabe}
R. Raabe et al., Phys. Rev. Lett. \textbf{101}, 212501 (2008).
%
\bibitem{Riisager}
K. Riisager et al., Phys. Lett. B 732, 305 (2014).
%
\bibitem{Tang18}
Z. Tang et al., arXiv:1802.01595 [nucl-ex], (2018).
%
\bibitem{Czar18}
A. Czarnecki, W.J. Marciano, A. Sirlin, arXiv:1802.01804 [hep-ph], (2018).
%
\bibitem{Sere18}
A.P. Serebrov et al., arXiv:1802.06277 [nucl-ex], (2018).
%
\bibitem{McKeen18}
D. McKeen et al., arXiv:1802.08244 [hep-ph], (2018).
%
\bibitem{Baym18}
G. Baym et al., arXiv:1802.0824482 [hep-ph], (2018).
%
\bibitem{Motta18}
T.F. Motta et al., arXiv:1802.08427 [nucl-th], (2018).
%
\end{thebibliography}
\end{document}